\documentclass[prl,twocolumn,floatfix,showpacs,superscriptaddress]{revtex4-1}
\usepackage{graphicx}
\usepackage{amsmath}
\begin{document}

\title{Mechanical annealing of metallic electrodes at the atomic scale}

\author{C. Sabater}
\affiliation
{Departamento  de  F\'{\i}sica Aplicada,
Universidad de Alicante, San Vicente del Raspeig, E-03690 Alicante, Spain}

\author{C. Untiedt}
\affiliation
{Departamento  de  F\'{\i}sica Aplicada,
Universidad de Alicante, San Vicente del Raspeig, E-03690 Alicante, Spain}

\author{J. J. Palacios}
\affiliation
{Departamento de F\'{\i}sica de la Materia Condensada, 
Universidad Aut\'{o}noma de Madrid, Madrid E-28049, Spain}

\author{M.J. Caturla}
\email{mj.caturla@ua.es}
\affiliation
{Departamento  de  F\'{\i}sica Aplicada,
Universidad de Alicante, San Vicente del Raspeig, E-03690 Alicante, Spain}

\date{\today}

\begin{abstract}

The process of creating an atomically defined and
robust metallic tip is described and quantified using measurements of contact conductance between gold
electrodes and numerical simulations. Our experiments show how the same conductance behavior can be obtained 
for hundreds of cycles of formation and rupture of the nanocontact 
by limiting the indentation depth between the two electrodes 
up to a conductance value of approximately $5G_{0}$ in the case of gold. This phenomenon is rationalized using 
molecular dynamics simulations together with density functional theory transport calculations which show how, 
after repeated indentations (mechanical annealing), the two metallic electrodes are shaped into tips of reproducible structure. 
These results provide a crucial insight into fundamental aspects relevant to
nano-tribology or scanning probe microscopies. 

\end{abstract}

\maketitle

An important source of uncertainty in electrical measurements at the atomic scale
is rooted in the indefinite structure of the metallic electrodes, regardless of whether they are 
created by mechanical, electrical, or electrochemical means\cite{Tsong90,Mel90}.  
Many technological applications and fundamental studies would benefit from atomically defined, stable, and reproducible metallic
electrodes down to the nanoscale.  For example, often before a measurement using scanning tunnelling 
microscopy (STM), the tip has to be prepared by indenting several times on a surface until a clear image can be obtained. 
To date, despite the importance of this process, there is no rigorous study or quantification of the underlying
physical mechanisms.   

The formation of atomic scale contacts between two metallic systems is usually performed by 
STM \cite{AgrRod93,PasMen93} or by mechanically controlled break junctions (MCBJ) \cite{MulRui92}.
The evolution of the minimum cross section of these contacts can be followed by recording traces of 
conductance while the electrodes are displaced with respect to each other. 
Generally, the conductance traces obtained differ between different experimental runs and 
a statistical approach is used, recording conductance histograms in 
order to identify preferential configurations \cite{OleLag95,KraRui95}. This phenomenon has been extensively studied  
both experimentally and theoretically \cite{AgrLev03}.

There are, however, certain situations where the same conductance traces are recurrent during hundreds of cycles of 
compression or traction. It was shown \cite{UntRub97}
that when a 100 atom thick contact is properly trained through repeated indentation, 
it is possible to improve its crystallinity and aspect-ratio as long as the wire 
is not broken during this process. 
In the opposite limit, Trouwborts et al \cite{TroHui08} showed that they could form stable one-atom contacts
by gently cycling the formation of a one-atom contact from rupture. These results have been recently modelled through molecular dynamics simulations \cite{Wang11}. 

In this paper we present, using both experiments and simulations,
a comprehensive study of the formation process of an atomically defined 
stable gold tip through repeated indentations. Experiments allow us to 
quantify the indentation values for the formation of a robust tip, while electronic and atomic structure
simulations provide an explanation  to this fact.
Remarkably, 
we show that reproducible traces can be obtained even when the tips are indented up to a conductance of $\approx 5G_{0}$, 
where plastic deformation was expected to occur. The results presented 
here shed light on the process of preparation of STM tips and, in general, 
on the formation of stable and reproducible metallic electrodes.

To produce atomic contacts we have used a low temperature STM where tip and sample
consist of two gold electrodes of purity above 99.995\% .
Prior to the experiments the samples are carefully cleaned in an acetone bath. All
experiments are performed at low temperatures (4.2K) under a cryogenic vacuum atmosphere, 
to ensure cleanness.
Applying a constant bias voltage of $V_{bias}= 100mV$ to the junction 
the electronic current is measured while displacing the two electrodes with respect each other.
In this way, conductance vs. electrode displacement is constructed during
the formation and rupture of the contact, reflecting its structural evolution. 
We have done a systematic study of the evolution and reproducibility of these traces,
comparing their behaviour for different values of the maximum conductance during indentation, $G_{max}$. 

Figure 1(a) shows 286 consecutive conductance traces during the formation  and rupture of gold nanocontacts when we limit the maximum 
value of conductance to $G_{max} = 5G_{0}$. The inset shows a three dimensional graph of the rupture cases. 
One can clearly observe that the traces remain 
unchanged during hundreds of cycles with most of them showing the same height and length for the different steps. 
This is a remarkable result considering that 
these conductance values correspond to cross sections of a few atoms where one would expect plastical deformation to occur.
In Figure 1(b) the conductance traces for formation and rupture of the gold contact are 
shown when the maximum conductance is limited to a value of 
$G_{max} = 8G_{0}$. The inset figure shows the three dimensional plot of the same traces. 
The difference between the traces in Fig. 1(a) and (b) are outstanding: for the larger
indentation the traces are not reproduced for consecutive cycles.

\begin{figure}[t]
\begin{center}
\includegraphics[scale=0.18]{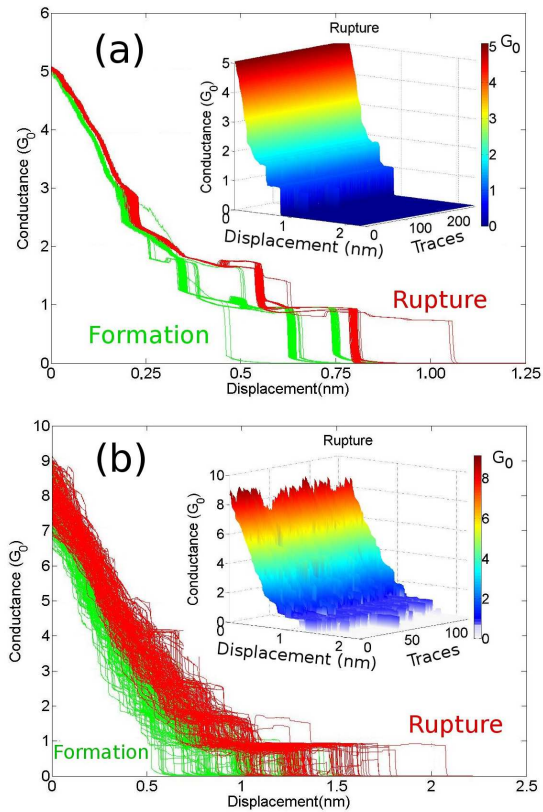}
\caption{Experimental traces obtained for Au nanocontacts during formation and rupture when limiting the conductance to (a) $5G_{0}$ and (b) $8G_{0}$.  
The inset shows a 3D figure of the rupture where the third axis is each individual trace} 
\end{center}
\label{figure:au.experiment}
\end{figure}

Two major conclusions can be drawn from these experimental results. 
On one hand, if one is interested in obtaining
a representative histogram of the conductance behavior of a particular metal,  exploring many different configurations,
it is important to indent the two electrodes to values higher than a threshold conductance  
($\approx 5G_{0}$ in the case of gold). Otherwise, one might end up in a stable 
configuration that will bias the statistical result. 
On the other hand, these results also show a quantitative way to obtain a robust STM tip or electrode: one must 
indent the tip several times to a conductance value not higher than a given threshold value. 
The values of conductance provided here are for the case of gold, but experiments in other materials are currently being performed. 

These experiments, however, do not provide atomic scale information or the mechanism 
that gives rise to this behaviour. Numerical simulations are now used in order to gain some 
understanding and provide an explanation consistent with the experimental 
results. Molecular dynamics (MD) with empirical potentials is the computational tool most commonly used to study 
the breaking and formation process of a nanocontact \cite{Landman,Sorensen98, HasMed01,GarPel05,Palacios05, DreHeu04, PauDre06,  CalCat08}.
The conductance is expected to be proportional to the minimum cross section of the contact in an MD simulation. Therefore, histograms of minimum cross sections can be correlated, more or less accurately, 
with conductance histograms. Usually this correlation is better done through other computational methods 
such as density functional theory (DFT) \cite{Palacios05} or tight-binding \cite{DreHeu04, PauDre06}  electronic transport
calculations. In these MD simulations different traces are obtained varying slightly the initial conditions, 
generally changing the initial velocity distributions, but using 
the same initial geometry. 

We have performed MD simulations of breaking and formation of gold
nanocontacts using the embedded atom potential developed by Zou et al.
 \cite{PotentialGold}. This potential is fitted to reproduce basic materials properties such as lattice constant, 
elastic constant, bulk moduli, and vacancy formation energy. 
The initial configuration is a neck, following the work of Sorensen et al \cite{Sorensen98}. 
The ratio between the length
of the neck and the narrowest cross section is 5 for simulations with 525 atoms and 2 for simulations with 2804 atoms. 
In this way we mimic a long and narrow nanowire and a short and wide constriction.
Three atomic layers at the top and at the bottom of the simulation cell are displaced by a fixed distance ( every time step, 
with characteristic strain rates between $10^{8}$ and $10^{10} s^{-1}$\cite{Sorensen98}. 
The temperature in the simulations is controlled in two different ways: it is kept constant by scaling the velocities of all 
atoms every time step (every femtosecond) as done in other works \cite{HasMed01, GarPel05, PauDre06}, or the temperature is not 
fixed during the calculation but it is scaled in every cycle, after the contact is broken. In this last case the maximum raise 
in temperature is 70K.
The initial temperature for both cases is 4.2 K. Unlike previous 
works \cite{HasMed01, GarPel05, DreHeu04, PauDre06, CalCat08}, calculations are performed in cycles, 
following the experimental procedure. 
A tension along the $[001]$ direction is applied to the initial structure until it breaks. 
Computing the minimum cross section \cite{Brat} of the nanocontact while stretching allows us to 
identify when then contact is broken. Once the contact is broken, the system is allowed to evolve 
during another 200 steps to ensure that the two tips are well separated. Then the two tips are brought in contact again (by applying a compression)
until a given cross section (which has varied) is reached. This cycle is repeated at least 20 times for each case.

Figure 2, top section, shows 6 snapshots of the nanocontact during the formation and breaking process 
for the case of the narrow nanowire. The results presented are the initial 
configuration and the configuration of the system before cycles 2, 5, 10, 15 and 20. 
In this particular calculation the temperature was not kept fixed 
and the two nanoelectrodes are brought together until the minimum cross section reaches a size of 15 atoms. 
The atoms have been colored according to
the following procedure: after the nanocontact is broken for the first time, atoms in the top nanoelectrode 
and in the bottom one are labeled differently, as shown in the configuration for cycle 2. Therefore, the number
of atoms that are exchanged between electrodes, with respect to the original configuration, can be followed (see cycles 5 to 20).
Figure 2 shows the percentage of atoms on one nanoelectrode that were originally on the second one, for the two nanoelectrodes, 
as a function of the indentation cycle. The figure shows how this number reaches a constant value after 7 - 8 cycles. Note that 
during the second cycle the nanocontact breaks from a different point leaving only one nanoelectrode with mixed
atoms. This can be seen more clearly in the movie of the complete calculation \cite{Movie1}.

\begin{figure}[t]
\includegraphics[scale=0.35]{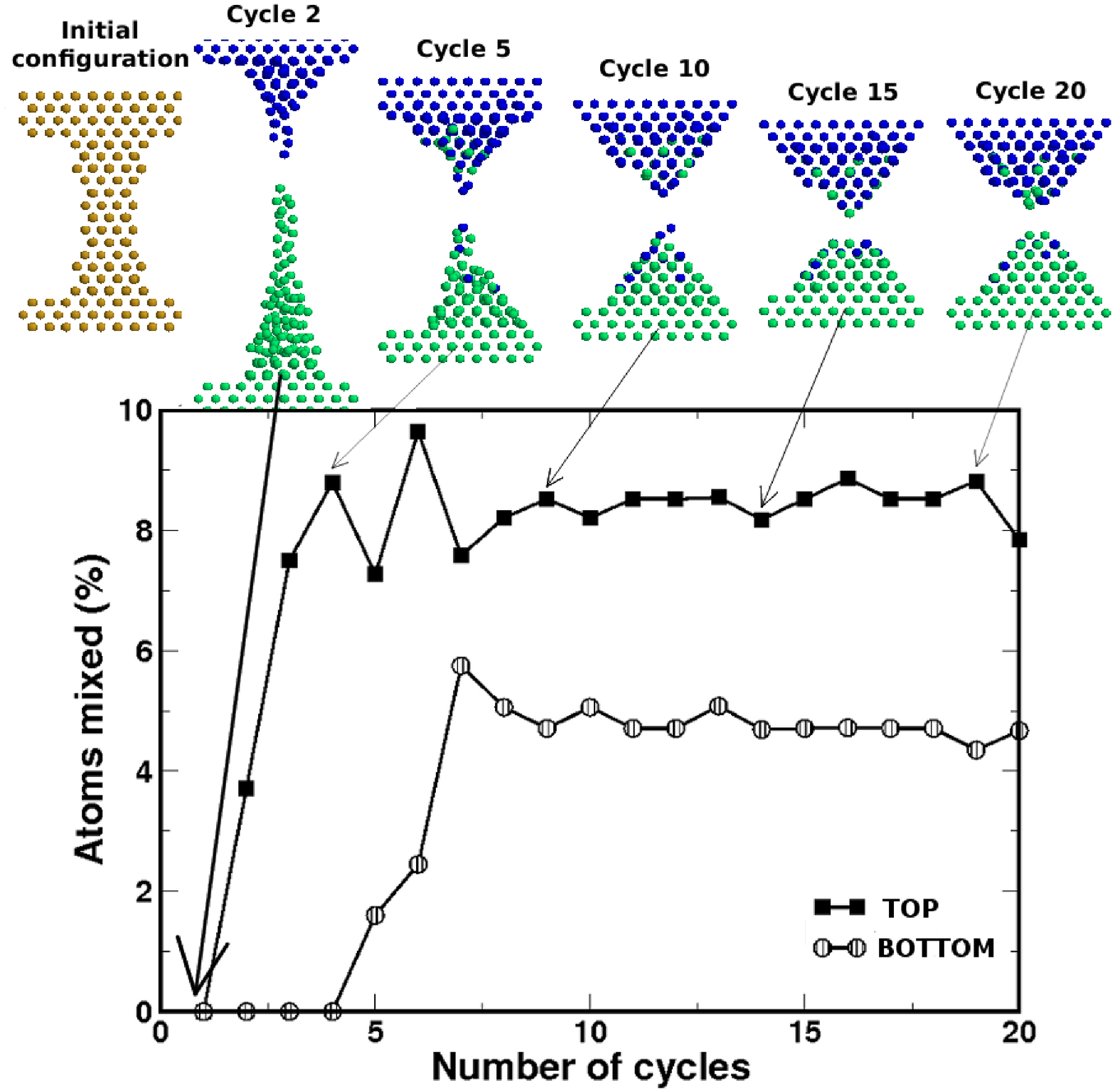}
\vspace{0.0cm}
\caption{Snapshots of the MD simulations of rupture and formation of a nanocontact in gold for the initial configuration and before cycles 2, 5, 10, 15 and 20 (top).
Number of atoms in the top nanoelectrode (in \%) that were initially on the second one, and viceversa, as a function of the number of cycles (bottom). Temperature was not fixed in this calculation.}
\label{figure:md.snapshots}
\end{figure}

The first interesting feature that can be observed from the different configurations in Figure 2 
is the re-shaping of the two sides with the indentation cycles.
The second cycle shows the formation of long chains, in this case a double chain, which 
have been already identified and described by other authors \cite{Double_ch}. However, 
as the indentation process continues, stable tips are formed which remain almost unchanged for cycles 10, 15 and 20. The constant
value in the atomic mixing between tips for these cycles also points to the formation of a stable configuration. 
The tips formed have a pyramidal shape with $(111)$ faces, which is consistent
with energetic considerations.

The formation of a robust tip after repeated indentations can also be observed in the traces of the minimum cross section
(in number of atoms) obtained from the simulations.
In figure 3 traces for all 20 indentations for the case of stretching the nanocontact until rupture are presented. 
It is clear from this figure that while the first 10 traces are very different from each other, 
the last 20 traces are quite similar, just like the experimental data shown in Fig. 1.

\begin{figure}[t]
\includegraphics[scale=0.30]{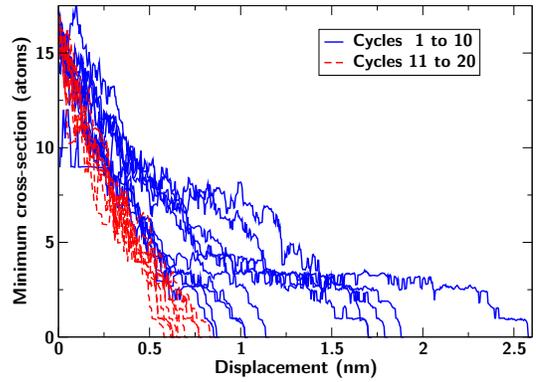}
\caption{Minimum cross section traces obtained from MD simulations for rupture of the nanocontact. Blue lines show the first 10 traces while red lines are the last 20 traces. 
Temperature was not fixed in this calculation.}
\label{figure:figure_3}
\end{figure}

The reproducibility of the traces in the simulations depends, at least, on two factors. On one hand, if the 
temperature is kept fixed in the simulations 
as done in previous calculations \cite{Sorensen98,HasMed01}, traces are much more similar to each 
other than if no heat dissipation is included. However, 
there is also less relaxation of the atoms in the two tips and that results, in some cases, in the formation
of unlikely structures. Note that keeping the temperature fixed is
equivalent to an almost instant dissipation of heat (femtoseconds). A second factor that influences the
reproducibility of the traces is the initial atomic configuration. When the initial structure
is a short and wide constriction, the two pyramidal tips are easily formed after the first indentation and all subsequent traces 
are very similar as long as the maximum indentation value is kept low. 
Figure 4 shows the minimum cross section for the case of 2804 atoms,
with fixed temperature and for three different indentation values: cross section of 5, 15 and 25 atoms. 
The inset in this figure shows the initial configuration in this simulation.
This result clearly shows that if the indentation is kept low, below a minimum cross section of 15 atoms, 
the traces are nicely reproduced, just like in the experiments shown in Fig. 1(a). 
However, when the indentation is up to a section of 25 atoms, the traces are very different 
from one cycle to the next, even in the case where heat is dissipated. This is also 
in agreement with the experimental data shown in figure 1(b). This behavior is the result of the
sliding between the two tips. For small indentations the tips can accomodate each other by simply
sliding along the $(111)$ faces with almost no disorder of any of the two tips. However, as the indentation 
proceeds disorder of the tips increases, changing their shapes and consequently the traces obtained (see \cite{movies2and3} for
movies of these calculations.

\begin{figure}[t]
\includegraphics[scale=0.3]{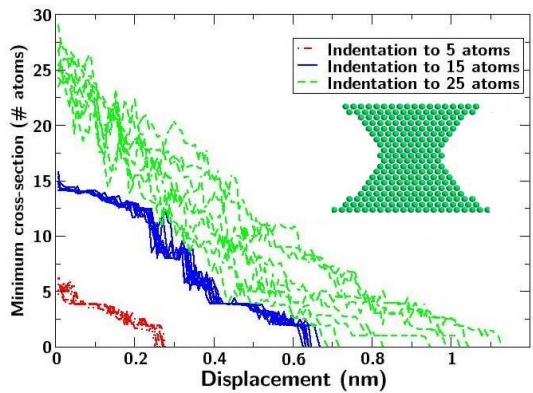}
\caption{Minimum cross section traces obtained from MD simulations for rupture of the nanocontact when indentation is followed up to a cross section of 5 atoms, 15 atoms
and 25 atoms. The inset shows the initial configuration of the contact with 2804 atoms. Temperature was fixed at every time step in these calculations.}
\label{figure:figure_4}
\end{figure}

MD simulations, however, do not provide precise information about conductance of these systems 
and therefore cannot be directly compared with the experimental measurements.
For the case of gold, which has a single $s$ valence orbital half occupied, 
one could use a simple rule of thumb and consider that the cross section in atoms is equal to the conductance
in units of $G_0$. However, this might not be the case 
when the minimum cross section has several atoms and the lattice is disordered. 
Therefore, we have performed DFT electronic transport calculations 
to obtain the conductance values for the atomic 
configurations obtained with the MD simulations. We have used our code ANT.G, which is part of 
the package ALACANT \cite{ALACANT} and implements the non-equilibrium Green's function
formalism with the help of the popular code GAUSSIAN09 \cite{GAUSSIAN09}.  

Figure 5 shows two conductance traces for the first and tenth ruptures along with the minimum cross section 
for the case of the narrow wire (525 atoms). 
Here, due to the large number of atoms, we have
employed a very basic basis set consisting only of one $6s$ orbital and one electron, the remaining 78 
electrons being part of the pseudopotential. For reassurance, we have also computed a few MD snapshots with a larger basis set 
consisting of $5s5p5d6s$ orbitals and 11 electrons (diamonds, in figure 5(a)). 
These calculations show that for small cross sections, up to $\approx 3-4 G_{0}$,
there is an equivalence between conductance and number of atoms in the minimum cross section. 
However, for larger cross sections this correspondence does no longer holds, which can be attributed, in part,
to the disordered structure. In this particular 
case the conductance of a cross section with $\approx 12$ atoms is reduced to only $\approx 6-8 G_{0}$. 
This could explain why in the molecular dynamics simulations described 
above we obtain stable traces for indentation up to cross sections of $\approx 15$ atoms, 
while experimentally the indentation must be limited to $\approx 5G_{0}$. 

\begin{figure}[t]
\includegraphics[scale=0.06]{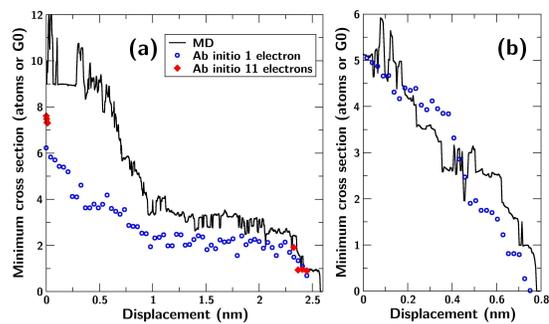}
\caption{Traces of conductance from DFT (open circles are calculations with 1 electron and diamonds are calculations with 11 electrons) and estimates from MD minimum cross section (lines) for calculations with 525 atoms. (a) Rupture trace during first cycle and (b) rupture trace for cycle number 10 for a maximum indentation of 5 atoms in cross section.}
\label{figure:md.abinitio}
\end{figure}

In summary, our experiments show that it is possible to create stable and atomically well-defined tips by repeated 
indentation of an electrode into a surface as long as the indentation is limited to a few conductance quanta: $\approx 5G_{0}$
in the case of gold.  Our simulations show how, under repeated indentations, two pyramidal tips with $(111)$
faces form. When the indentation is small, the tips will slide on each other producing almost no distortion and, therefore,
reproducible traces. As the indentation increases so does the disorder and the reproducibility disappears. 

This work was supported by the Spanish government through grants FIS2010-21883 and
CONSOLIDER CSD2007-0010.

\end{document}